\documentstyle[aps,twocolumn]{revtex}

\renewcommand{\vec}[1]{\bbox{#1}}
\begin{document}
\draft
\title{Universality of the Tangential Shape Exponent at the Facet Edge
 of a Crystal}\author{Yasuhiro Akutsu}
\address{Department of Physics, Graduate School of Science, Osaka University,
Toyonaka 560, Osaka, Japan}
\author{Noriko Akutsu}
\address{Faculty of Engineering, Osaka Electro-Communication University,
 Neyagawa 572, Osaka, Japan}
\author{Takao Yamamoto}
\address{Department of Physics, Faculty of Engineering, Gunma University,
 Kiryu 376, Gunma, Japan}
\date{March 13, 1998}
\maketitle
\begin{abstract}
Below the roughening temperature, the equilibrium crystal shape (ECS)
 is composed of both facets and a smoothly curved surface. As for the
 ``normal'' profile (perpendicular to the facet contour), the ECS has
 the exponent 3/2 which is characteristic of systems in the Gruber-Mullins-Pokrovsky-Talapov
 universality class.  Quite recently, it was pointed out that the ECS
 have a ``new'' exponent 3 for ``tangential'' profile.  We first show
 that this behavior is universal because it is a direct consequence of
 the well-establised universal form of the vicinal-surface free energy
 ($p$: surface gradient): $f(p)=f(0)+\gamma p +B p^3 +O(p^4)$.  Second,
 we give a universal relation between the amplitudes of the tangential
 and the normal profiles, in close connection with the universal Gaussian
 curvature jump at the facet edge in systems with short-range inter-step
 interactions.  Effects of the long-range interactions are briefly discussed.
\end{abstract}
%
%
%
%

\narrowtext

A crystal surface with small average tilt relative to a crystal axis is
 called  vicinal surface.  Below the roughening temperature $T_{{\rm R}}$
 the vicinal surface is well descibed by the terrace-step-kink (TSK) picture
 where the surface is made up of many ``flat'' terraces connected by steps.
  Since the step is a unidirectional linear object, the vicinal surface,
 as an assembly of steps, belongs to the Gruber-Mullins-Pokrovsky-Talapov
 (GMPT) universality class.\cite{1}  In this view, the vicinal-surface
 free energy $f(\rho)$ (per projected area) as a function of the step
 density $\rho$ has the well-known form of the expansion,\cite{1,2} $f(\rho)=f(0)+a_1\rho+a_3\rho^3+O(\rho^4)$.
  This form of surface free energy leads to, via the Wulff construction,\cite{3}
 the universal exponent 3/2 of the equilibrium crystal shape (ECS, for
 short) near the facet edge, $z\sim (\Delta x)^{3/2}$, where we have chosen
 the $z$ axis to be facet normal, and $\Delta x$ to be the ``normal''
 distance from a facet edge.\cite{4,4a}  Careful consideration of the
 crystal anisotropy gives us a universal relation\cite{5} between the
 coefficients $a_{1}$ and $a_{3}$.  This relation leads to the universal
 Gaussian curvature jump at the facet edge\cite{5,6,7} whose physical
 origin is, therefore, quite different\cite{8} from the universal curvature
 jump at $T_{{\rm R}}$.\cite{4}

Very recently, Dahmen {\em et al}.\cite{9} pointed out that, if we approach
 the facet edge ``tangentially'' (i.e., along the tangential direction
 at the facet edge), ECS behaves as $z\sim (\Delta y)^3$  where $\Delta
 y$ is the distance from the facet edge along the tangential direction.
  They drew this conclusion from an exact calculation (essentially equivalent
 to Ref.8) on the body-centered-cubic solid-on-solid model\cite{10} (BCSOS
 model, for short).  It is an interesting problem, then, to see whether
 this ``new'' exponent is universal or not.  Also, relations, if any,
 to the above-listed universal behaviors are worth to be explored.  To
 clarify these points is the aim of the present paper.  As will be seen
 below, the tangential exponent 3 is a direct consequence of the GMPT-form
 of expansion of the surface free energy, hence, is universal.  Further,
 the universal relation between the expansion coefficients leads also
 to a universal relation between the amplitudes of the normal and the
 tangential ECS profiles.

Consider a facet and its neighboring curved surface of the ECS with macroscopic
 size at a temperature $T$ below $T_{{\rm R}}$ (of the facet, to be precise).
  We use the Cartesian coordinates $(x,y,z)$ with $z$-direction chosen
 to be facet normal (downword, for convenience), which allows us to describe
 the ECS by an equation $z=z(x,y)$. The surface gradient at a position
 $(x,y,z(x,y))$ is conveniently expressed by the two-dimensional gradient
 vector $\vec{p}=(p_x, p_y)$ defined by
\begin{equation}
p_x=\frac{\partial z}{\partial x},\quad p_y=\frac{\partial z}{\partial
 y}.\label{tag1}
\end{equation}
By $f(\vec{p})$ we denote the free energy per projected area\cite{11}
 of the surface with fixed mean gradient $\vec{p}$.  Introducing a ``field''
  $\vec{\eta}$ conjugate to $\vec{p}$, we can perform the Legendre transformation
 $\vec{p}\rightarrow \vec{\eta}$, $f(\vec{p})\rightarrow \tilde{f}(\vec{\eta})$,
 with
\begin{equation}
\tilde{f}(\vec{\eta})=f(\vec{p})-\vec{\eta}\cdot\vec{p}. \label{tag2}
\end{equation}
The Andreev free energy $\tilde{f}(\vec{\eta})$ directly gives us the
 ECS as\cite{11}
\begin{equation}
z =-\frac{1}{\lambda}\tilde{f}(\lambda \vec{r}),\label{tag3}
\end{equation}
where $\vec{r}=(x,y)$  and $\lambda$ is a scale factor.  In what follows
 we consider ``normalized'' ECS  by setting $\lambda=1$ in (\ref{tag3}).
  Then the field $\vec{\eta}$ is just the two-dimensional position vector
 of the ECS.

Properties of the ECS near the facet edge are governed by the small-$|\vec{p}|$
 behavior of $f(\vec{p})$ which has the GMPT-type form of expansion. 
 Properly taking account the crystal anisotropy, we have the expansion
 of the form,
\begin{equation}
f(\vec{p})=f(0)+\gamma(\theta)|\vec{p}|+B(\theta)|\vec{p}|^3 +O(|\vec{p}|^4),
 \label{tag4}
\end{equation}
where we have introduced the angle variable $\theta$ by 
\begin{equation}
p_{x}=|\vec{p}|\cos\theta,\quad p_{y}=|\vec{p}|\sin\theta. \label{tag5}
\end{equation}
Physically, $\gamma(\theta)$ is the step tension and $\theta$ the mean
 running direction angle of steps on the vicinal surface with gradient
 $\vec{p}$.  Note that $\theta$ is also the direction angle of the tangential
 line of the facet contour (Fig.1).   In systems with only short-range
 inter-step interactions, the coefficient $B(\theta)$ is always positive
 and is universally given by\cite{5}
\begin{equation}
B(\theta)=\frac{\pi^2(k_{{\rm B}}T)^2}{6\tilde{\gamma}(\theta)}, \label{tag6}
\end{equation}
where $\tilde{\gamma}(\theta)=\gamma(\theta)+\partial^2\gamma(\theta)/\partial\theta^2$
 is the step stiffness.  In association with the Legendre transformation,
 we have the following relation between the two-dimensional position vector
 $\vec{r}=(x,y)$ and the gradient $\vec{p}$:
\begin{equation}
\vec{r}=\frac{\partial f(\vec{p})}{\partial \vec{p}}. \label{tag7}
\end{equation}
Near the facet edge in the curved region, we obtain from (\ref{tag4})
 (neglecting $O(|\vec{p}|^{4})$ and higher-order terms)
\begin{eqnarray}
x&=&x_c(\theta)+|\vec{p}|^2[3B(\theta)\cos\theta-B'(\theta)\sin\theta]
 \label{tag8a},\\
y&=&y_c(\theta)+|\vec{p}|^2[3B(\theta)\sin\theta+B'(\theta)\cos\theta]
 ,\label{tag8b}
\end{eqnarray}
where the curve $\{(x_{c}(\theta),y_{c}(\theta))\}_{\theta}$ is the facet
 contour corresponding to the zero-gradient limit of the curved region.
  Explicitly, we have
\begin{eqnarray}
x_c(\theta)&=&\gamma(\theta)\cos\theta -\gamma'(\theta)\sin\theta ,\label{tag9a}\\\
y_c(\theta)&=&\gamma(\theta)\sin\theta +\gamma'(\theta)\cos\theta .\label{tag9b}
\end{eqnarray}
We should note that (\ref{tag9a}) and (\ref{tag9b}) are precisely the
 equations determining the {\em two-dimensional} ECS (=facet shape) from
 $\gamma(\theta)$ regarded as a one-dimensional interface tension.\cite{12}

Let us now discuss the ECS near the facet edge in detail.  For a given
 ECS described by an equation $z=z(x,y)$, we specify any point on the
 ECS surface by the two-dimensional position vector $(x,y)$.  For convenience,
 we take the $xy$ plane (i.e., $z=0$) to be the facet plane, which corresponds
 to putting $f(0)=0$ in (\ref{tag4}).  Fix a point $P$ on the facet contour,
 and choose the $x$ and $y$ axes so that the $y$-axis is parallel to the
 tangential line of the facet contour at $P$, and $x$-axis perpendicular
 to it (Fig.2).  With this choice of the coordinate system, we have $\theta=0$
 at $P$ and $\gamma'(0)=0$ (see (\ref{tag5}) and (\ref{tag9b})).  Our
 task is to obtain the ECS profile close to the point $P=(x_c(0),y_c(0))=(x_c(0),0)$.

Along the $x$-axis ($\theta=0$, $\Delta x\equiv x-x_c(0)$), we have from
 (\ref{tag8a}) and (\ref{tag8b})
\begin{equation}
|\vec{p}|=p_x=\frac{1}{\sqrt{3B(0)}} (\Delta x)^{1/2}, \label{tag10}
\end{equation}
giving the ``normal'' profile with the well-known exponent $\theta_x=3/2$:
\begin{equation}
 z \sim \frac{2}{3\sqrt{3B(0)}}(\Delta x)^{3/2} .  \label{tag11}
\end{equation}
The profile (\ref{tag11}) leads to the divergent behavior of the normal
 curvature $\kappa_x\approx \partial^2 z/\partial x^2 \sim (\Delta x)^{-1/2}$
 near the facet edge.  In the light of the universal $Gaussian$ curvature
 jump at the facet edge,\cite{5} the ``tangential curvature'' $\kappa_y
 \approx \partial^2 z/\partial y^2$, {\em along the $x$ axis}, vanishes\cite{5,13}
 as $(\Delta x)^{1/2}$ [Gaussian curvature is a product of two pricipal
 curvatures, $\kappa_{x}$ and $\kappa_{y}$]. We should note that a different
 exponent $\theta_y$ for the tangential profile $z\sim (\Delta y)^{\theta_y}$
 ($\Delta y=y-y_c(0)$) with $\theta_y>2$ has already been implied by the
 vanishing of the tangential curvature at $P$.

The actual value of $\theta_y$ can be derived as follows.  Note that,
 along the tangential line ($x-x_c(0)=0$), $\theta$ and $|\vec{p}|$ are
 not independent but are constrained to satisfy, 
\begin{equation}
-\frac{1}{2}\tilde{\gamma}(0)\theta^2+3B(0)|\vec{p}|^2 =0,\label{tag12}
\end{equation}
which can be derived by expanding (\ref{tag8a}) and (\ref{tag9a}) with
 respect to $\theta$ and $|\vec{p}|$ ($|\vec{p}|<<1$ and $|\theta|<<1$,
 very near the point $P$).  Combining (\ref{tag12}) with (\ref{tag8a})--(\ref{tag9b}),
 we obtain
\begin{eqnarray}
\theta &=&\frac{\Delta y}{\tilde{\gamma}(0)},\label{tag13a}\\
|\vec{p}|&=&\frac{\Delta y}{\sqrt{6\tilde{\gamma}(0)B(0)}},\label{tag13b}
\end{eqnarray}
along the $y$-direction. Near $P$, $-z=\tilde{f}(x_c(0),\Delta y+y_c(0))$
 is expanded to give
\begin{eqnarray}
-z &=& [\gamma(0)+ \frac{1}{2}\gamma''(0)\theta^2]|\vec{p}| + B(0)|\vec{p}|^3
 \nonumber \\
   & &- \Delta y |\vec{p}|\theta -x_c(0)|\vec{p}|(1-\frac{1}{2}\theta^2).
 \label{tag14}
\end{eqnarray}
Putting (\ref{tag13a}) and (\ref{tag13b}) into (\ref{tag14}), we have
\begin{equation}
z=\frac{1}{3\sqrt{6B(0)\tilde{\gamma}(0)^3}}(\Delta y)^3 , \label{tag15}
\end{equation}
giving $\theta_y=3$. We should remark here that the results (\ref{tag11})
 and (\ref{tag15}) apply to any point on the facet contour, because the
 point $P$ has been chosen arbitrarily; hence, $\tilde{\gamma}(0)$ and
 $B(0)$ in (\ref{tag11}) and (\ref{tag15}) can be replaced by $\tilde{\gamma}(\theta)$
 and $B(\theta)$.

We have another simple geometrical derivation of (\ref{tag15}) as follows.
 Very near the point $P$, we take a different point $Q$ on the facet contour
 (Fig.3). Along the $x'$-direction chosen normal to the facet contour
 at $Q$, ECS profile has the form (\ref{tag11}) with $B(0)$ replaced by
 $B(\theta)$ where $\theta$ ($|\theta|<<1$) is the tangent angle (relative
 to $y$-axis at $P$) of the facet contour at $Q$. Note that the facet
 contour near $P$ is approximately a part of a circle with its radius
 being the curvature radius $R$ which is proportional to the step stiffness
 $\tilde{\gamma}(0)$.\cite{14}  For the normalized ECS we have $R=\tilde{\gamma}(0)$,
 hence, by an elementary geometry we can relate $\Delta x'$ (distance
 along the $x'$-direction) to $\Delta y$ as
\begin{equation}
\Delta x' = \frac{(\Delta y)^2}{2R} = \frac{(\Delta y)^2}{2\tilde{\gamma}(0)}.
 \label{tag16}
\end{equation}
Replacing this $\Delta x'$ with $\Delta x$ in (\ref{tag11}) we reproduce
 (\ref{tag15}) [we can put $B(\theta)\approx B(0)$, for our purpose].

We thus have shown that the ``new exponent'' $\theta_y=3$ is a direct
 consequense of the well-established GMPT-type expansion of the vicinal
 surface free energy (\ref{tag4}).  Let us next discuss the ``critical
 amplitudes'' of the profiles.  In the derivation of the universal Gaussian
 curvature jump at the facet edge,\cite{5} the universal relation
\begin{equation}
B(\theta)=\frac{\pi^2(k_{{\rm B}}T)^2}{6\tilde{\gamma}(\theta)}, \label{tag17}
\end{equation}
which holds for any system (with only short-range inter-step interactions)
 is essential. This relation was originally derived in the coarse-grained
 TSK picture of the vicinal surface,\cite{5} and has been confirmed in
 several ways.\cite{5,6,7,15,15a}  For example, exact calculation for
 the BCSOS model have verified (\ref{tag17}) for arbitrary $\theta$.\cite{7}
 Using (\ref{tag17}), we obtain a universal relation between the amplitudes
 of the normal and tangential profiles as follows.  By $A_x(\theta)$ and
 $A_y(\theta)$ we denote the amplitudes of the ECS profiles, namely,
\begin{eqnarray}
z \sim \left\{
\begin{array}{ll}
A_x(\theta) (\Delta x)^{3/2}&\mbox{(normal direction)},\\
A_{y}(\theta) (\Delta y)^3 &\mbox{(tangential direction)}\\
\end{array}
\right. . \label{tag18}
\end{eqnarray}
Restoring the scale factor $\lambda$ in (\ref{tag3}), we have from (\ref{tag11}),
 (\ref{tag15}) and (\ref{tag17}), 
\begin{equation}
[A_{x}^2(\theta)A_{y}(\theta)]^{1/3}=\lambda\frac{2}{3\pi k_{{\rm B}}T},
 \label{tag19}
\end{equation}
which means that at a fixed temperature, the quantity  $(A_x^2A_y)^{1/3}$
 is constant along the facet contour.  The scale factor $\lambda$ can
 be determined, for example, from the measurable ratio $\kappa/\sigma(\theta)^2$
 ($\propto \lambda/(k_{{\rm B}}T)$\cite{14}) where $\kappa$ is the curvature
 of the facet contour and $\sigma$ the scaled fluctuation width of a single
 step.\cite{16} 

Before closing, we give a brief comment on the effect of the long-range
 inter-step interaction ($\sim 1/r^2$, with $r$ being the ter-step distance)
 which has its origin mainly in the elastic deformation, and  is important
 in discussing real crystal surfaces.\cite{17}  It has been known that
 inclusion of $g/r^2$ interaction with positive coupling constant $g$
 does not modify the GMPT-type form of expansion, but merely renormalizes
 the coefficient $B$.  Explicit form of the renomalized $B$ has also been
 known.\cite{15,18,19}  Hence the exponent $\theta_y=3$ does not change
 with the $g/r^2$ interaction.  Since $\gamma(\theta)$ is a quantity associated
 with a single isolated step and is not affected by the $g/r^2$ interaction,
 the relation between $\gamma$ and the renormalized $B$ should be modified.
 If $g$ is $\theta$-independent ($\theta$: mean running direction of steps)
 then the universal relation (\ref{tag19}) still holds in a modified form.
  However, if $g$ depends on $\theta$, $A_{x}$ and $A_{y}$ will no longer
 be universally related to each other.  In this case, (\ref{tag19}) may
 provide a way to determine the $\theta$-dependence of $g$ experimentally,
 by measuring the ratio $[A_{x}^2(\theta)A_{y}(\theta)]^{1/3}/[A_{x}^2(0)A_{y}(0)]^{1/3}$.

To summarize, we have discussed the tangential profile of the equilibrium
 crystal shape near the facet edge below the roughening temperature. We
 have shown that the profile $z\sim (\Delta y)^{\theta_y}$ ($\Delta y$:
 distance along the tangential direction at the facet edge) with $\theta_y=3$
 is a direct consequence of the Gruber-Mullins-Pokrovsky-Talapov type
 expansion of the surface free energy $f(\vec{p})=f(0)+\gamma(\theta)|\vec{p}|+B(\theta)|\vec{p}|^3+\ldots$,
 which implies that the exponent 3 is universal.  Further, we have presented
 a general relation between the amplitudes of the normal and the tangential
 profiles, which results from the known universal relation between the
 coefficient $B(\theta)$ and the step stiffness $\tilde{\gamma}(\theta)=\gamma(\theta)+\gamma''(\theta)$.

 This work was partially supported by  the ``Research for the Future"
 Program  from The Japan Society for the Promotion of Science (JSPS-RFTF97P00201)
 and by the Grant-in-Aid for Scientific Research from Ministry of Education,
 Science, Sports and Culture (No.09640462).

\newpage
\begin{flushleft}
{\bf Figure Captions}
\end{flushleft}

\noindent
{\bf Fig.1}:  Facet and vicinal surface. Upper right is an ``atomic scale''
 view of the surface near the facet edge in the curved region, which can
 be regarded as an assembly of steps forming a vicinal surface.

\noindent
{\bf Fig.2}:  Choice of $x$- and $y$-axes at $P$ on the facet contour.

\noindent
{\bf Fig.3}:  Geometrical derivation of Eq.(18).  For $Q$ very near $P$,
 ``normal distance'' $\Delta x'$  relates to ``tangential distance'' $\Delta
 y$  as $\Delta x'=(\Delta y)^2/2R$, where $R$ is the curvature radius
 at $P$.

\end{document}